\documentclass[preprint,aps ,nofootinbib]{revtex4}
\pdfoutput=1
\usepackage{graphicx}
\usepackage{epsfig}
\usepackage{amsmath}
\usepackage{amsfonts}
\usepackage{amssymb}
\usepackage{color}%
\usepackage{dcolumn}
\providecommand{\U}[1]{\protect\rule{.1in}{.1in}}

\newcommand{\f}{\begin{equation}}
\newcommand{\ff}{\end{equation}}
\newcommand{\fa}{\begin{eqnarray}}
\newcommand{\ffa}{\end{eqnarray}}

\begin{document}
\title{Holographic Fermionic Liquid with Lattices}
\author{Yi Ling $^{1,2}$}
\email{lingy@ihep.ac.cn}
\author{Chao Niu $^{1}$}
\email{niuc@ihep.ac.cn}
\author{Jian-Pin Wu $^{3,4,5}$} \email{jianpinwu@yahoo.com.cn}
\author{Zhuo-Yu Xian $^{6,1}$}
\email{x.zhuoyu@mail.scut.edu.cn}
\author{Hongbao Zhang $^{7,8}$} \email{hzhang@vub.ac.be}
\affiliation{ $^1$ Institute of High Energy Physics, Chinese
Academy of Sciences, Beijing 100049, China\\
$^2$  State Key Laboratory of Theoretical Physics, Institute of
Theoretical Physics, Chinese Academy of Sciences, Beijing 100190,
China\\
$^3$Department of Physics, Hanyang University, Seoul 133-791, Korea\\
$^4$Center for Quantum Spacetime, Sogang University, Seoul 121-742, Korea\\
$^5$Department of Physics, School of Mathematics and Physics,
Bohai University,
Jinzhou 121013, China\\ $^6$ Department of Physics, South China University of Technology, Guangzhou 510641, China\\
$^7$Theoretische Natuurkunde, Vrije Universiteit Brussel, \\
Pleinlaan 2, B-1050 Brussels, Belgium\\
$^8$The International Solvay Institutes,\\
Pleinlaan 2, B-1050 Brussels, Belgium }

\begin{abstract}
We investigate the holographic fermions over a gravitational
lattice background with a rather low temperature. Since the
rotation symmetry is broken on the plane, the lattice effects
change the shape of the Fermi surface within the first Brillouin
zone from a circle to an ellipse. When the Fermi surface
intersects with the Brillouin zone boundary, the band structure
with a band gap is observed through a numerical analysis. We
construct a lattice model sourced by a scalar field as well as an
ionic lattice model without the scalar field. In both cases we
find the similar physical results.

\end{abstract} \maketitle

\section{Introduction}
As a powerful tool of understanding a variety of strongly coupled
condensed matter systems, AdS/CMT has been entering an era from
building purely theoretical models with highly symmetric
configuration to modelling more realistic condensed matter system
with less symmetry. In particular, instead of working within the
probe limit, very recently the authors in \cite{Horowitz:2012ky} and
\cite{Horowitz:2012gs} have constructed some spatially inhomogeneous but
periodic gravitational backgrounds by fully solving the coupled
partial differential equations numerically with the
Einstein-DeTurck method. Such gravitational backgrounds, by
holography, correspond to the boundary systems in the presence of
a lattice, a key ingredient in the condensed matter systems. Thus
with such gravitational backgrounds, one can explore various
lattice effects in the holographic investigation of condensed
matter systems. Actually for the transport coefficients such as
optical conductivity and thermoelectric conductivity, it has been
shown that not only does the presence of a lattice result in the
Drude peak at low frequencies but also induces a new intermediate
scaling regime in which a robust power law behavior is found with
respect to the frequency\cite{Horowitz:2012ky,Horowitz:2012gs}. Remarkably such a result
is in striking agreement with the experiments on the cuprates,
including the superconducting phase\cite{Horowitz:2013jaa}.

The purpose of this paper is to investigate how such a holographic
lattice affects the Fermi surface by putting the Dirac field in
such bulk gravitational backgrounds. To achieve this, in the next
section we first build up the holographic framework to extract the
spectral function for fermions living on the boundary by solving
the bulk Dirac equation in periodic backgrounds. Then we shall
present a numerical construction of two kinds of ultra cold
holographic lattices using the Einstein-DeTurck method in Section
III. After that, the numerical results for the Fermi surface in
the presence of such holographic lattices are detailed in Section
IV. We shall conclude with some discussions as well as further
directions in the final section.

\section {Holographic setup for the bulk Dirac field in periodic backgrounds}

Start from the bulk action for a Dirac field with mass $m$ and charge $q$
\begin{eqnarray}
\label{actionspinor}
S_{D}=i\int d^{4}x \sqrt{-g}\overline{\zeta}\left(\Gamma^{a}\mathcal{D}_{a}-m\right)\zeta,
\end{eqnarray}

in the static background as general as
\begin{eqnarray}
\label{MetricAnsatz}
ds^{2}&=&-g_{tt}(x,z)dt^{2}+g_{zz}(x,z)dz^{2}+g_{xx}(x,z)dx^{2}+g_{yy}(x,z)dy^{2}+2g_{xz}(x,z)dxdz,\nonumber\\
A&=&A_t(x,z)dt.
\end{eqnarray}

Here $\Gamma^{a}=(e_{\mu})^{a}\Gamma^{\mu}$ with $(e_{\mu})^{a}$ a set of orthogonal normal
vector bases and $\Gamma^{\mu}$ Gamma matrices. In addition, $\mathcal{D}_{a}$ is the covariant derivative, i.e.,
\begin{eqnarray}
\label{Dderivative}
\mathcal{D}_{a}=\partial_{a}+\frac{1}{4}(\omega_{\mu\nu})_{a}\Gamma^{\mu\nu}-iqA_{a},
\end{eqnarray}
where $\Gamma^{\mu\nu}=\frac{1}{2}[\Gamma^{\mu},\Gamma^{\nu}]$, and $(\omega_{\mu\nu})_{a}$ are the spin connection 1-forms given by
\begin{eqnarray}
\label{spinconnectionD}
(\omega_{\mu\nu})_{a}=(e_{\mu})_{b}\nabla_{a}(e_{\nu})^{b}.
\end{eqnarray}

With the following orthogonal normal vector bases
\begin{eqnarray}
\label{VectorBases}
&&
(e_{0})^{a}=\frac{1}{\sqrt{g_{tt}}}(\frac{\partial}{\partial t})^{a},~~
(e_{1})^{a}=\frac{1}{\sqrt{g_{xx}}}(\frac{\partial}{\partial x})^{a},~~
(e_{2})^{a}=\frac{1}{\sqrt{g_{yy}}}(\frac{\partial}{\partial y})^{a},~~
\nonumber\\
&&
(e_{3})^{a}=-\sqrt{\frac{g_{xx}}{g_{xx}g_{zz}-g_{xz}^{2}}}(\frac{\partial}{\partial z})^{a}
+\frac{g_{xz}}{\sqrt{g_{xx}(g_{xx}g_{zz}-g_{xz}^{2})}}(\frac{\partial}{\partial x})^{a},
\end{eqnarray}
the non-vanishing
components of spin connections can be calculated as follows
\begin{eqnarray}
\label{SpinConnections}
(\omega_{01})_{a}=-(\omega_{10})_{a}&=&-\frac{\partial_{x}g_{tt}}{2\sqrt{g_{tt}g_{xx}}}(dt)_{a},
\nonumber\\
(\omega_{03})_{a}=-(\omega_{30})_{a}&=&\frac{g_{xx}\partial_{z}g_{tt}-g_{xz}\partial_{x}g_{tt}}
{2\sqrt{g_{tt}g_{xx}(g_{xx}g_{zz}-g_{xz}^{2})}}(dt)_{a},
\nonumber\\
(\omega_{12})_{a}=-(\omega_{21})_{a}&=&-\frac{\partial_{x}g_{yy}}{2\sqrt{g_{xx}g_{yy}}}(dy)_{a},
\nonumber\\
(\omega_{13})_{a}=-(\omega_{31})_{a}&=&\left(-\frac{\partial_{z}g_{xx}}{2\sqrt{g_{xx}g_{zz}-g_{xz}^{2}}}
+\frac{2g_{xx}\partial_{x}g_{xz}-g_{xz}\partial_{x}g_{xx}}{2g_{xx}\sqrt{g_{xx}g_{zz}-g_{xz}^{2}}}\right)(dx)_{a}
\nonumber\\
&&
+\frac{g_{xx}\partial_{x}g_{zz}-g_{xz}\partial_{z}g_{xx}}{2g_{xx}\sqrt{g_{xx}g_{zz}-g_{xz}^{2}}}(dz)_{a},
\nonumber\\
(\omega_{23})_{a}=-(\omega_{32})_{a}&=&\frac{g_{xz}\partial_{x}g_{yy}-g_{xx}\partial_{z}g_{yy}}
{2\sqrt{g_{xx}g_{yy}(g_{xx}g_{zz}-g_{xz}^{2})}}(dy)_{a}.
\end{eqnarray}

Thus the Dirac equation
\begin{eqnarray}
\label{DiracEquation1}
\Gamma^{a}\mathcal{D}_{a}\zeta-m\zeta=0
\end{eqnarray}
can be written as
\begin{eqnarray}
\label{DiracESimplify}
&&
-\sqrt{\frac{g_{xx}}{g_{xx}g_{zz}-g_{xz}^{2}}}\Gamma^{3}\partial_{z}\zeta
+\frac{1}{\sqrt{g_{tt}}}\Gamma^{0}(\partial_{t}-i q A_{t})\zeta
+\left( \frac{1}{\sqrt{g_{xx}}}\Gamma^{1} + \frac{g_{xz}}{\sqrt{g_{xx}(g_{xx}g_{zz}-g_{xz}^{2})}} \Gamma^{3} \right)\partial_{x}\zeta
\nonumber\\
&&
+\frac{1}{\sqrt{g_{yy}}}\Gamma^{2}\partial_{y}\zeta
+\left(\frac{\partial_{x}g_{tt}}{4g_{tt}\sqrt{g_{xx}}}
+\frac{\partial_{x}g_{yy}}{4g_{yy}\sqrt{g_{xx}}}
+\frac{\sqrt{g_{xx}}\partial_{x}g_{zz}}{4(g_{xx}g_{zz}-g_{xz}^{2})}
+\frac{g_{xz}(g_{xz}\partial_{x}g_{xx}-2g_{xx}\partial_{x}g_{xz})}{4g_{xx}^{3/2}(g_{xx}g_{zz}-g_{xz}^{2})}\right)
\nonumber\\
&&
\times\Gamma^{1}\zeta-\frac{1}{4\sqrt{g_{xx}(g_{xx}g_{zz}-g_{xz}^{2})}}\times
\nonumber\\
&&
\left(\partial_{z}g_{xx}
-2\partial_{x}g_{xz}
+\frac{g_{xz}}{g_{xx}}\partial_{x}g_{xx}
+\frac{g_{xx}\partial_{z}g_{tt}-g_{xz}\partial_{x}g_{tt}}{g_{tt}}+\frac{g_{xx}\partial_{z}g_{yy}-g_{xz}\partial_{x}g_{yy}}{g_{yy}}\right)
\Gamma^{3}\zeta-m\zeta=0.\nonumber\\
\end{eqnarray}
To proceed, let us first make a transformation $\zeta=(g_{tt}g_{xx}g_{yy})^{-\frac{1}{4}}\mathcal{F}$.  Then
 the above equation turns out to be
\begin{eqnarray}
\label{DiracEF}
&&
-\sqrt{\frac{g_{xx}}{g_{xx}g_{zz}-g_{xz}^{2}}}\Gamma^{3}\partial_{z}\mathcal{F}
+\frac{1}{\sqrt{g_{tt}}}\Gamma^{0}(\partial_{t}-i q A_{t})\mathcal{F}
+\left( \frac{1}{\sqrt{g_{xx}}}\Gamma^{1} + \frac{g_{xz}}{\sqrt{g_{xx}(g_{xx}g_{zz}-g_{xz}^{2})}} \Gamma^{3} \right)\partial_{x}\mathcal{F}
\nonumber\\
&&
+\frac{1}{\sqrt{g_{yy}}}\Gamma^{2}\partial_{y}\mathcal{F}
+\left(-\frac{\partial_{x}g_{xx}}{4g_{xx}^{3/2}}
+\frac{\sqrt{g_{xx}}\partial_{x}g_{zz}}{4(g_{xx}g_{zz}-g_{xz}^{2})}
+\frac{g_{xz}(g_{xz}\partial_{x}g_{xx}-2g_{xx}\partial_{x}g_{xz})}{4g_{xx}^{3/2}(g_{xx}g_{zz}-g_{xz}^{2})}\right)
\Gamma^{1}\mathcal{F}
\nonumber\\
&&
+\frac{1}{4\sqrt{g_{xx}(g_{xx}g_{zz}-g_{xz}^{2})}}
\left(
2\partial_{x}g_{xz}
-2\frac{g_{xz}}{g_{xx}}\partial_{x}g_{xx}
\right)
\Gamma^{3}\mathcal{F}
-m\mathcal{F}
=0
\end{eqnarray}
Next expanding $\mathcal{F}$ as $\mathcal{F}=F(x,z) e^{-i\omega t +ik_{i}x^{i}}$,
one can have
\begin{eqnarray}
\label{DiracEFourier}
\Delta_{3}\Gamma^{3}F+\Delta_{0}\Gamma^{0}F-\Delta_{1}\Gamma^{1}F-\Delta_{2}\Gamma^{2}F+mF=0,
\end{eqnarray}
where we have denoted
\begin{eqnarray}
\label{DiracEDenote}
&&
\Delta_{3}=:\frac{1}{\sqrt{g_{xx}(g_{xx}g_{zz}-g_{xz}^{2})}}
\left(g_{xx}\partial_{z} - g_{xz}\partial_{x} - ik_{1}g_{xz}
-\frac{1}{2}\partial_{x}g_{xz} + \frac{g_{xz}}{2g_{xx}}\partial_{x}g_{xx} \right),
\nonumber\\
&&
\Delta_{0}=:i(\omega+ q A_{t})\frac{1}{\sqrt{g_{tt}}},
\nonumber\\
&&
\Delta_{1}=:\left(\frac{1}{\sqrt{g_{xx}}}\partial_{x}
+ \frac{ik_{1}}{\sqrt{g_{xx}}}
-\frac{\partial_{x}g_{xx}}{4g_{xx}^{3/2}}
+\frac{\sqrt{g_{xx}}\partial_{x}g_{zz}}{4(g_{xx}g_{zz}-g_{xz}^{2})}
+\frac{g_{xz}(g_{xz}\partial_{x}g_{xx}-2g_{xx}\partial_{x}g_{xz})}{4g_{xx}^{3/2}(g_{xx}g_{zz}-g_{xz}^{2})}\right),
\nonumber\\
&&
\Delta_{2}=:\frac{ik_{2}}{\sqrt{g_{yy}}}.
\end{eqnarray}
Now if we choose our gamma matrices as
\begin{eqnarray}
\label{GammaMatrices}
 && \Gamma^{3} = \left( \begin{array}{cc}
-\sigma^3  & 0  \\
0 & -\sigma^3
\end{array} \right), \;\;
 \Gamma^{0} = \left( \begin{array}{cc}
 i \sigma^1  & 0  \\
0 & i \sigma^1
\end{array} \right),  \cr
&&
\Gamma^{1} = \left( \begin{array}{cc}
-\sigma^2  & 0  \\
0 & \sigma^2
\end{array} \right), \;\;
 \Gamma^{2} = \left( \begin{array}{cc}
 0  & \sigma^2  \\
\sigma^2 & 0
\end{array} \right),
\end{eqnarray}
and split the 4-component spinor into two 2-component spinors as $F=(F_{1},F_{2})^{T}$, then
the Dirac equation becomes
\begin{eqnarray}
\label{DiracEF1F2}
\Delta_{3}
\left( \begin{matrix} F_{1} \cr  F_{2} \end{matrix}\right)
-m\sigma^3\otimes\left( \begin{matrix} F_{1} \cr  F_{2} \end{matrix}\right)
+\Delta_{0}\sigma^2\otimes\left( \begin{matrix} F_{1} \cr  F_{2} \end{matrix}\right)
\pm i\Delta_{1}
\sigma^1 \otimes \left( \begin{matrix} F_{1} \cr  F_{2} \end{matrix}\right)
- i\Delta_{2} \sigma^1\otimes \left( \begin{matrix} F_{2} \cr  F_{1} \end{matrix}\right)
=0.
\end{eqnarray}
Note that this is a coupled equation between $F_{1}$ and $F_{2}$ as opposed to the case in \cite{Faulkner:2009wj}, where $F_1$ and $F_2$ are decoupled.
Furthermore, by the decomposition
\begin{eqnarray} \label{gammarDecompose}
F_{\alpha} \equiv \left( \begin{matrix} \mathcal{A}_{\alpha} \cr  \mathcal{B}_{\alpha} \end{matrix}\right)
\end{eqnarray}
with $\alpha=1,2$, the Dirac equation (\ref{DiracEF1F2}) can be expressed as
\begin{eqnarray} \label{DiracEAB1R}
(\Delta_{30}\partial_{z}+\Delta_{31}\mp m)
\left( \begin{matrix} \mathcal{A}_{1} \cr  \mathcal{B}_{1} \end{matrix}\right)
\mp i\Delta_{0}\left( \begin{matrix} \mathcal{B}_{1} \cr  \mathcal{A}_{1} \end{matrix}\right)
+ i\Delta_{1} \left( \begin{matrix} \mathcal{B}_{1} \cr  \mathcal{A}_{1} \end{matrix}\right)
- i\Delta_{2} \left( \begin{matrix} \mathcal{B}_{2} \cr  \mathcal{A}_{2} \end{matrix}\right)
=0,
\end{eqnarray}
\begin{eqnarray} \label{DiracEAB2R}
(\Delta_{30}\partial_{z}+\Delta_{31}\mp m)
\left( \begin{matrix} \mathcal{A}_{2} \cr  \mathcal{B}_{2} \end{matrix}\right)
\mp i\Delta_{0}\left( \begin{matrix} \mathcal{B}_{2} \cr  \mathcal{A}_{2} \end{matrix}\right)
- i\Delta_{1} \left( \begin{matrix} \mathcal{B}_{2} \cr  \mathcal{A}_{2} \end{matrix}\right)
- i\Delta_{2} \left( \begin{matrix} \mathcal{B}_{1} \cr  \mathcal{A}_{1} \end{matrix}\right)
=0,
\end{eqnarray}
where
\begin{eqnarray} \label{Delta3031}
&&
\Delta_{30}=:\frac{g_{xx}}{\sqrt{g_{xx}(g_{xx}g_{zz}-g_{xz}^{2})}},
\nonumber\\
&&
\Delta_{31}=:\frac{- ik_{1}g_{xz}-\frac{1}{2}\partial_{x}g_{xz} + \frac{g_{xz}}{2g_{xx}}\partial_{x}g_{xx}}
{\sqrt{g_{xx}(g_{xx}g_{zz}-g_{xz}^{2})}}.
\end{eqnarray}

Suppose our background field is periodic along the $x$ direction
with the periodicity $c$. Then by the Bloch theorem, the solution
to our Dirac equation can always be expanded as follows
\begin{equation}\label{Bloch}
\left( \begin{matrix} \mathcal{A}_{\alpha}(x,z) \cr  \mathcal{B}_{\alpha}(x,z) \end{matrix}\right)=\sum_{n=0,\pm 1,\pm 2,\cdot\cdot\cdot}\left( \begin{matrix} \mathcal{A}_{\alpha,n}(z) \cr  \mathcal{B}_{\alpha,n}(z) \end{matrix}\right)e^{inKx}\end{equation}
with $K=\frac{2\pi}{c}$.

Below we will restrict ourselves to the
following periodic background, i.e.,
\begin{eqnarray}
\label{MetricConcrete}
ds^{2}&=&\frac{1}{z^{2}}\{-(1-z)P(z)Q_{tt}(x,z)dt^{2}
+\frac{Q_{zz}(x,z)dz^{2}}{P(z)(1-z)}\nonumber\\
&&+Q_{xx}(x,z)[dx+z^{2}Q_{xz}(x,z)dz]^{2}
+Q_{yy}(x,z)dy^{2}\},\nonumber\\
A&=&(1-z)\psi(x,z)dt.
\end{eqnarray}
Here
\begin{eqnarray}
\label{Pz}
P(z)=1+z+z^{2}-\frac{\mu_{1}^{2}z^{3}}{2}
\end{eqnarray}
with the other variables regular from the horizon $z=1$ all the
way to the conformal boundary $z=0$. In particular, we require at
the horizon
\begin{equation}
Q_{tt}(x,1)=Q_{zz}(x,1)
\end{equation}
and at the conformal boundary
\begin{equation}
Q_{tt}(x,0)=Q_{zz}(x,0)=Q_{xx}(x,0)=Q_{yy}(x,0)=1, Q_{xz}(x,0)=0, \psi(x,0)=\mu(x).
\end{equation}
Hence such a geometry asymptotes AdS with the curvature radius $L=1$ and the temperature given by
\begin{eqnarray}
T=\frac{P(1)}{4\pi}=\frac{6-\mu_{1}^{2}}{8\pi}.\label{temperature0}
\end{eqnarray}

With the above background ansatz, our Dirac equation gives rise to
\begin{eqnarray} \label{DiracEABalphaNearHorizon}
\partial_{z}\left( \begin{matrix} \mathcal{A}_{\alpha,n} \cr  \mathcal{B}_{\alpha,n} \end{matrix}\right)
\pm \frac{\omega}{4\pi T}\frac{1}{1-z}
\left( \begin{matrix} \mathcal{B}_{\alpha,n} \cr  \mathcal{A}_{\alpha,n} \end{matrix}\right)
=0
\end{eqnarray}
at the horizon. In order to obtain the retarded Green function on
the boundary by holography, the independent ingoing boundary
condition should be imposed at the horizon, i.e.,
\begin{equation}
\left( \begin{matrix} \mathcal{A}_{\alpha,n} \cr
\mathcal{B}_{\alpha,n} \end{matrix}\right)=\left( \begin{matrix} 1
\cr  -i\end{matrix}\right)(1-z)^{-\frac{i\omega}{4\pi
T}}\end{equation} for each $\alpha$ and $n$ with the others turned
off. On the other hand, near the AdS boundary, our Dirac equation
reduces to
\begin{eqnarray} \label{DiracEboundary}
(z\partial_{z}-m\sigma^3)\otimes \left( \begin{matrix} F_{1,n} \cr  F_{2,n} \end{matrix}\right)=0~.
\end{eqnarray}
Hence the solution can be asymptotically expanded near the AdS boundary as
\begin{eqnarray} \label{BoundaryBehaviour}
F_{\alpha,n} {\approx} a_{\alpha,n}z^{m}\left( \begin{matrix} 1 \cr  0 \end{matrix}\right)
+b_{\alpha,n}z^{-m}\left( \begin{matrix} 0 \cr 1 \end{matrix}\right).
\end{eqnarray}
Holography tells us that the retarded Green function can be obtained by the following relation
\begin{equation}
a_{\alpha,n}(\beta,l)=G_{\alpha,n;\alpha',n'}b_{\alpha',n'}(\beta,l),
\end{equation}
where $a_{\alpha,n}(\beta,l)$ and $b_{\alpha,n}(\beta,l)$ are the
asymptotic expansion coefficients in (\ref{BoundaryBehaviour}) of
the solution to the Dirac equation evolving from the ingoing boundary
condition with the only $(\beta,l)$ mode turned on.

In what follows, for simplicity but without loss of generality, we shall work solely with the case of $m=0$.

\section{Numerical construction of ultra cold holographic lattices}

A holographic lattice background can be constructed in at least
two ways. One way is to introduce a neutral scalar field with the
periodic boundary conditions along the spatial direction. We would
like to refer to this kind of holographic lattice as the scalar
lattice. The other kind of holographic lattice is induced directly
by a periodic chemical potential to the gauge field on the
boundary, which is referred to as the ionic lattice. This ionic lattice has been discussed earlier
but almost always treated perturbatively in \cite{Hartnoll:2012rj,Maeda:2011pk,Liu:2012tr,Flauger:2010tv,Hutasoit:2012ib}. In \cite{Horowitz:2012gs} it was treated exactly
without the scalar field. Both of these two kinds of
holographic lattice can be numerically constructed by the
Einstein-Deturck method, which has been detailed in
\cite{Horowitz:2012ky,Headrick:2009pv,Figueras:2011va}. In order to see the
lattice effect on the Fermi surface, we will follow the same route
to construct these two kinds of  holographic lattice backgrounds
with a rather cold temperature.

\subsection{The scalar lattice}

To construct an ultra cold scalar lattice, let us
consider the following gravitational action including a Maxwell field and a neutral scalar field, i.e.,
\begin{eqnarray} \label{GActionCold}
S=\frac{1}{16\pi G_{N}}\int d^{4}x \sqrt{-g}\left[R+\frac{6}{L^{2}}-\frac{1}{2}F_{ab}F^{ab}
-2\nabla_{a}\Phi\nabla^{a}\Phi-2M^{2}\Phi^{2}\right],
\end{eqnarray}
where $L$ is the AdS radius as before, and $M$ is the mass of the scalar field. In what follows, we shall set $L=1$ and $M^2=-2$.
Then the equations of motion can be derived from the above action as
\begin{eqnarray} \label{Delta3031}
&&
R_{ab}+3g_{ab}
-2\left(\nabla_{a}\Phi\nabla_{b}\Phi-\Phi^{2}g_{ab}\right)
-\left(F_{ac}F_{b}^{\ c}-\frac{g_{ab}}{4}F_{cd}F^{cd}\right)=0,
\ \\
&&
\nabla_{a}F^{a}_{\ b}=0,
\ \\
&&
\square\Phi +2\Phi=0.
\end{eqnarray}
Now with the ansatz in Eq.(\ref{MetricConcrete}) for our
holographic lattice, the asymptotic behavior of the scalar field is
\begin{equation}
\Phi =z\phi=z[\phi_1+z\phi_2+\mathcal{O}(z^2)]
\end{equation}
near the AdS boundary. Furthermore, the periodic structure can actually be induced by setting the source
\begin{equation}
\phi_1(x) =A_0cos(k_0x),
\end{equation}
and keeping the chemical potential fixed as
\begin{equation}
\mu(x)=\mu.
\end{equation}

\begin{figure}
\center{
\includegraphics[scale=0.275]{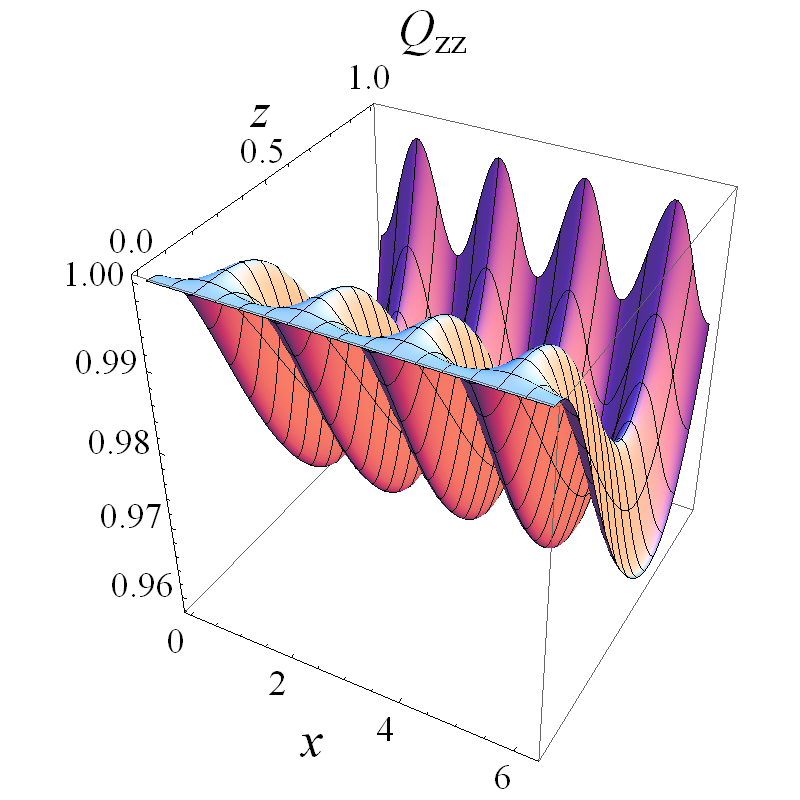}\hspace{0.1cm}
\includegraphics[scale=0.275]{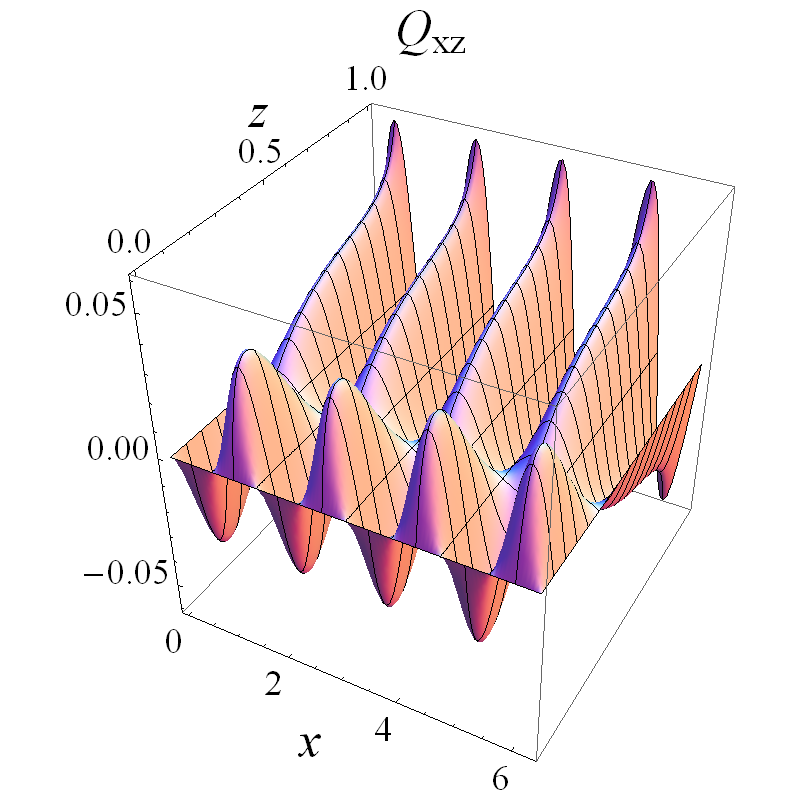}\\ \hspace{0.1cm}}
\center{
\includegraphics[scale=0.275]{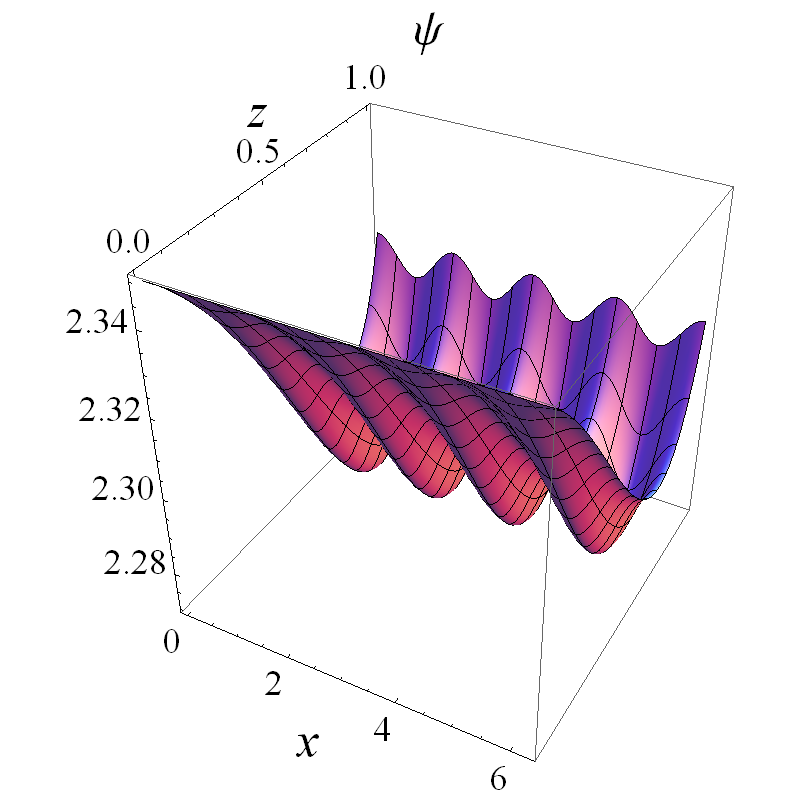}\hspace{0.1cm}
\includegraphics[scale=0.275]{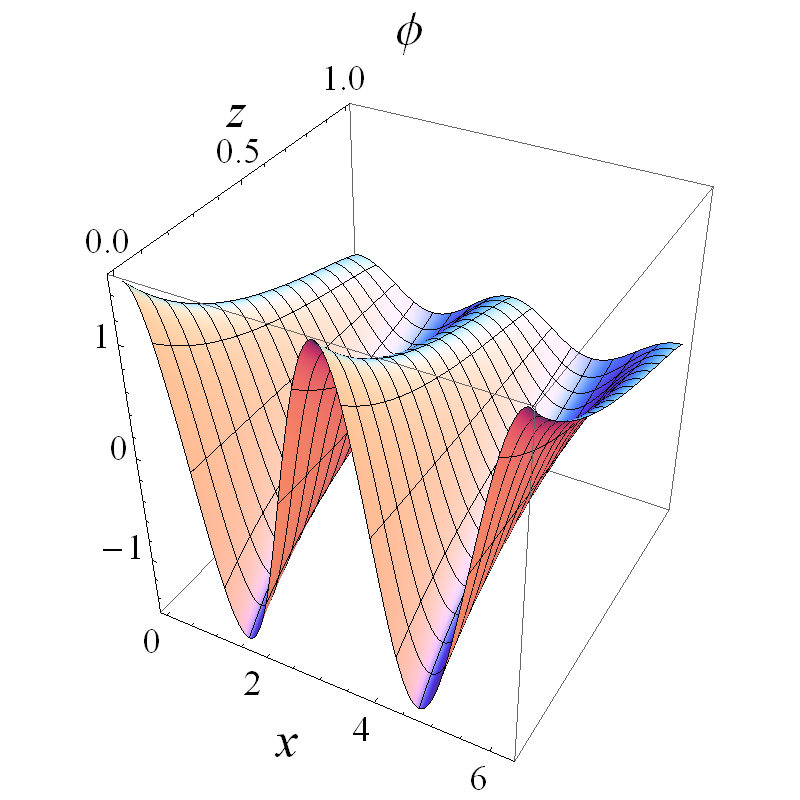}\\ \hspace{0.1cm}
\caption{\label{scalar}We show $Q_{zz}$, $Q_{xz}$, $\psi$ and
$\phi$ for $k_0=2$, $A_0 =1.5$, $\mu = 2.35$ and $T/\mu = 0.0081$.}
}
\end{figure}

For our purpose we would like to construct an ultra cold lattice
whose temperature is controlled by the parameter $\mu_1$ in
Eq.(\ref{temperature0}). This can be achieved by solving the
coupled Einstein-Maxwell-scalar equations numerically using the
Einstein-DeTurck method. Our numerical method is changing the
partial differential equations into non-linear algebraic equations
by the standard pseudospectral collocation approximation, and then
solving them by employing a Newton-Raphson method. As an example,
we show a solution to these equations in Fig.\ref{scalar} with
$A_0=1.5,k_0=2,\mu=\mu_1=2.35$, which corresponds  to a lattice
with $T/\mu= 0.0081$. Note that different from the scalar field
whose period is given by $2\pi/k_0$, the corresponding solutions
of all the components of the metric as well as the gauge field
have a period of $\pi/k_0$ along the $x$ direction, which comes
essentially from the fact that in our first equation of motion the
stress tensor is quadratic in the scalar field $\Phi$. Since the
scalar field does not appear in the Dirac equation, we claim that
the lattice constant that the probe Dirac field feel is $\pi/k_0$.
Therefore, the parameter $K$ in Eq.(\ref{Bloch}) is given by
$K=2k_0$.

\begin{figure}
\center{
\includegraphics[scale=0.3]{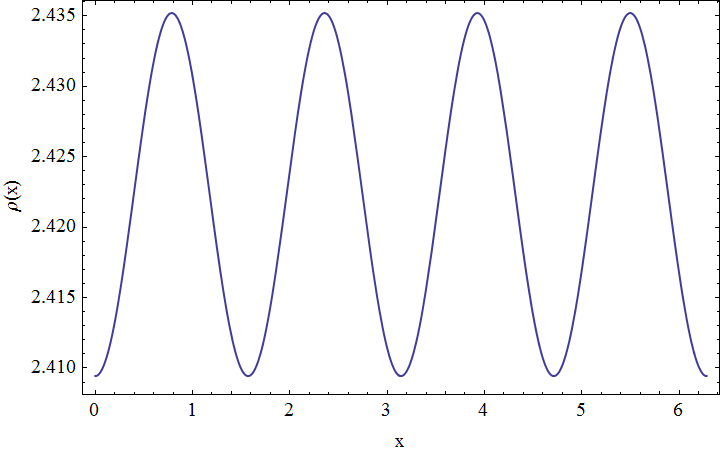}\hspace{0.1cm}
\includegraphics[scale=0.3]{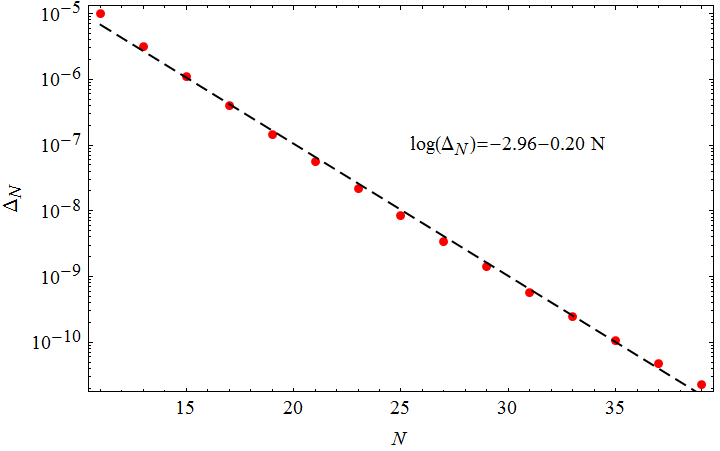}\\ \hspace{0.1cm}
\caption{\label{tendency}On the left we show the boundary charge density
$\rho$ as a function of $x$, which can be read off by expanding $\psi=\mu+(\mu-\rho)z+\mathcal{O}(z^2)$. On the right we show the charge discrepancy $\Delta_N$ as a function of the number of
grid points $N$, where the boundary charge is defined as $Q=\int_0^{2\pi/k_0}dx \rho$. The vertical scale is logarithmic, and the data
is well fit by an exponential decay: $\log(\Delta_N) =
-2.96-0.20\,N$.} }
\end{figure}

It is worthwhile to point out that during the course of numerical
analysis the lower the temperature is, the harder
monitoring the accuracy is. In order to guarantee the convergence
of our method at the given temperature, we are required to demonstrate the decaying
tendency of the charge discrepancy $\Delta_N$ which is defined
as $\Delta_N=|1-Q_N/Q_{N+1}|$ with $Q_N$  the charge on the
boundary when the number of grid points takes $N$. As shown in
in Fig.\ref{tendency}, such a decay is exponential, implying our results
are exponential convergence with the increasing of the grid
points. Moreover, we have also checked the behavior of the DeTurck
vector field $\xi^a$ which is defined in \cite{Horowitz:2012ky} and found that
for our solution $\xi^a\xi_a$ is smaller than $10^{-10}$,
ensuring that our numerics is leading to an
Einstein solution rather than a Ricci soliton.

\subsection{The ionic lattice}
\begin{figure}
\center{
\includegraphics[scale=0.275]{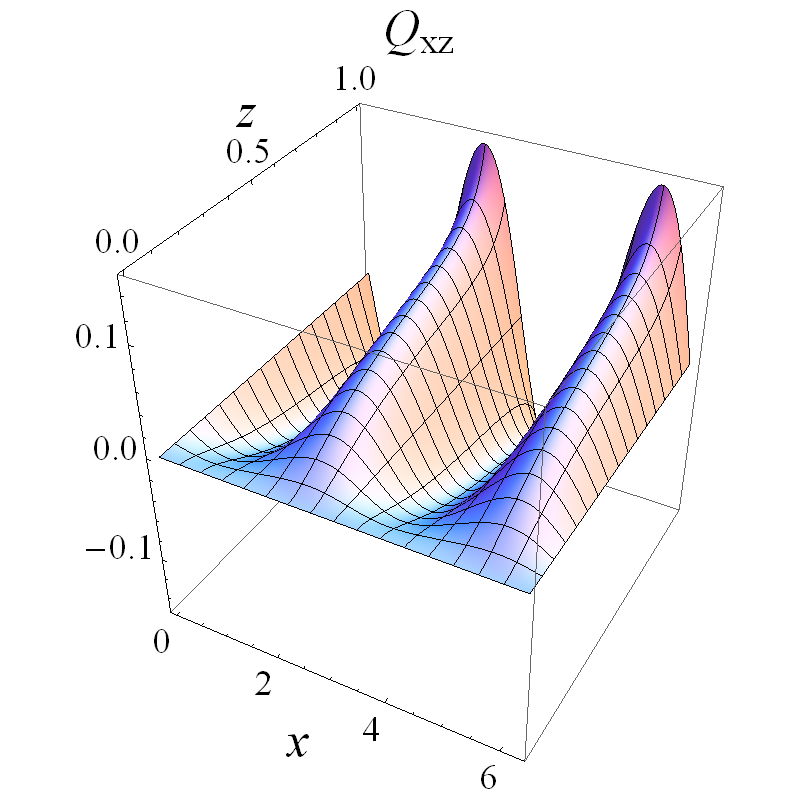}\hspace{0.1cm}
\includegraphics[scale=0.275]{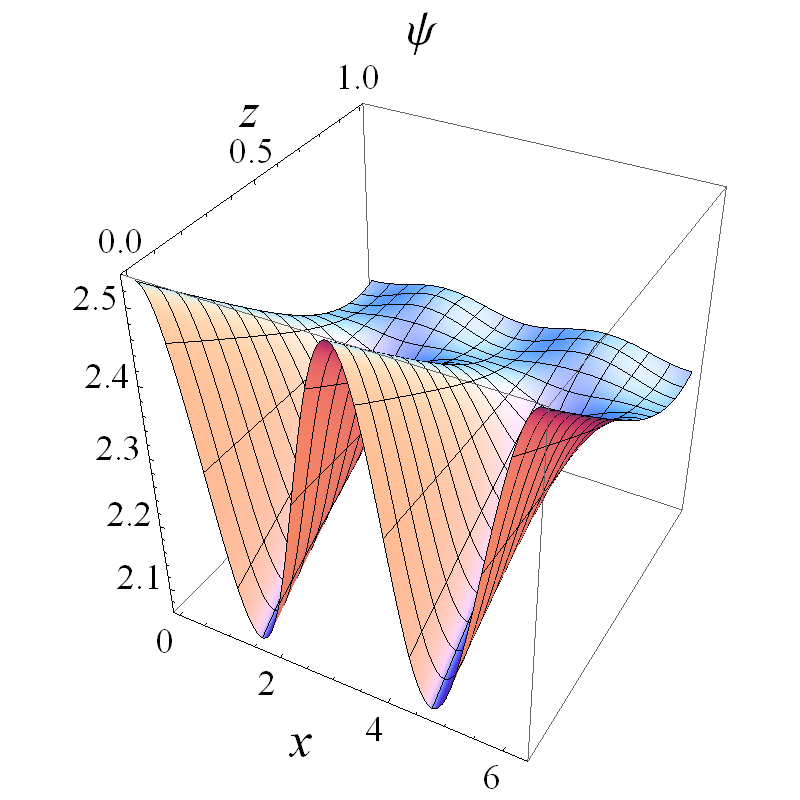}\\ \hspace{0.1cm}}
\center{
\includegraphics[scale=0.3]{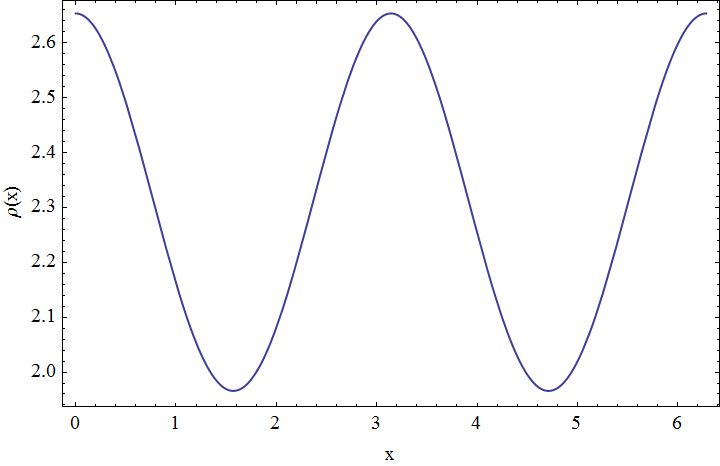} \hspace{0.1cm}
\includegraphics[scale=0.3]{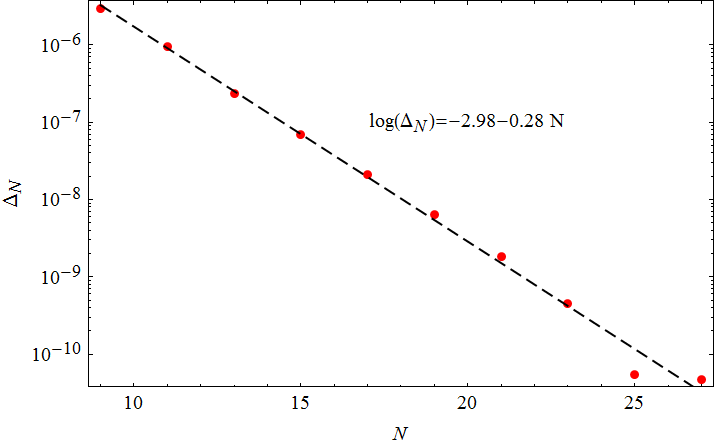}\\ \hspace{0.1cm}
\caption{\label{chemical}We show $Q_{xz}$ and
$\psi$ for $k_0=2$, $A_0 =0.1$, $\mu = 2.3$ and $T/\mu = 0.01$ in the top and the corresponding boundary charge density as well as the tendency of charge discrepancy in the bottom.}
}
\end{figure}

Compared to the scalar lattice, the ionic lattice can be generated simply by turning off the scalar field in (\ref{GActionCold}) and imposing the spatially varying boundary condition for the chemical potential as
\begin{equation}
\psi(x,0) =\mu[1+A_0cos(k_0x)].
\end{equation}
As an example, we illustrate a solution of $Q_{xz}$ and $\psi$ in
Fig.\ref{chemical} with $k_0=2$, $A_0 =0.1$, $\mu =\mu_1=2.3$,
which corresponds to $T/\mu = 0.01$. We stress that this type of
solutions are different from those for the scalar lattice. Namely
the chemical potential is periodic rather than a constant in the
scalar lattice. As a result, the boundary charge density in the
ionic lattice is expected to vary more dramatically than that in
the scalar lattice. This can easily be seen in our plot for the
charge density in Fig.\ref{chemical}. In this sense, one will also
expect that the effect onto the Fermi surface due to the ionic
lattice should be much stronger than that due to the scalar
lattice. We shall show in the subsequent section that this is
actually the case.

\section{Numerical results for the Fermi surface}

The existence of Fermi surfaces for holographic fermionic liquids
has been shown in various circumstances. We refer to
\cite{Faulkner:2011tm} and \cite{Iqbal:2011ae} for a recent review. In this paper we
focus on reporting the following two new results relevant to the
Fermi surface when a lattice is introduced for the gravitational
background. One is the shape of Fermi surface. In general one
expects that the shape of Fermi surface is not circular any more
since the rotation symmetry is broken in the presence of the
lattice along the $x$ direction. But how the Fermi surface would
change appears obscure since the Dirac equation becomes very
complicated in this case. Even in the case in which the lattice
effect can be treated by conventional condensed matter approach,
there is no some kind of universal result on the shape of Fermi
surface. Instead the shape will depend on the specific behavior of
periodic potential. Nevertheless to our surprise, our numerical
analysis gives us a very simple and elegant answer to this issue.
That is, when our holographic Fermi surface is located within the
first Brillouin zone, the shape is always an ellipse irrespective
of the specific value of parameters in question. The other is the
emergence of band gap at the intersection of the Fermi surface
with the Brillouin zone boundary. This phenomenon coincides
perfectly with the familiar lattice effect as one expects. Now let
us demonstrate our numerical results in detail.

To proceed, we would like to remark on how to identify the Fermi
surface in our current setting. Note that we are working with
holographic fermionic liquid in the presence of lattice at very
low but non-zero temperature, where the Fermi surface is somewhat
ill defined, because the concept of Fermi surface can only be
defined at zero temperature with translation symmetry unbroken.
However, as pointed out in \cite{Liu:2012tr}, the ARPES experiment is
actually blind to the lattice by smearing it into a continuum.
The measured spectral function can be captured by the
imaginary part of diagonal components of retarded Green function,
namely
$A(\omega,k_x=k_1+nK,k_y=k_2)=\mathbf{Im}(G_{1,n;1,n}+G_{2,n;2n})$.
Furthermore, taking into account that our ultra cold lattice only
smears the Fermi surface in a negligible way, we can locate the
position of Fermi surface by searching the peak of
$A(\omega,k_x,k_y)$ with the tiny frequency $\omega$ in the
momentum space. Such an identification of Fermi surface is also
described in \cite{Benini:2010qc} and similar to the operational definition
given in \cite{Kanigel}.

\begin{figure}
\center{
\includegraphics[scale=0.4]{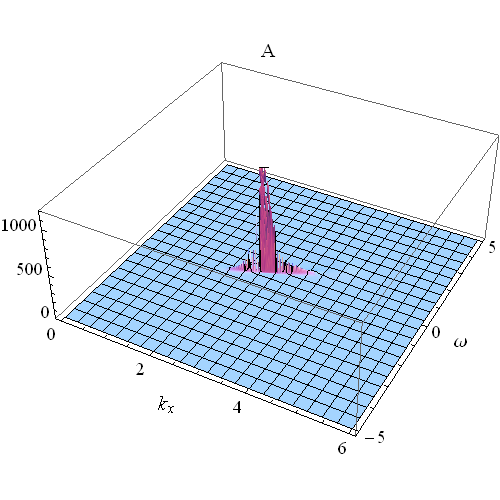}\hspace{0.1cm}
\caption{\label{FS1}We show the 3d plot of $A(\omega,k_x,0)$ for
$k_0=2$, $A_0 =1.5$, $\mu = 2.35$, $T/\mu = 0.0081$ and $q=1.7$. A
sharp peak near $\omega=0$ implies a Fermi surface with the Fermi
momentum $k_F=2.634$.} }
\end{figure}

\begin{figure}[h]
\center{
\includegraphics[scale=0.6]{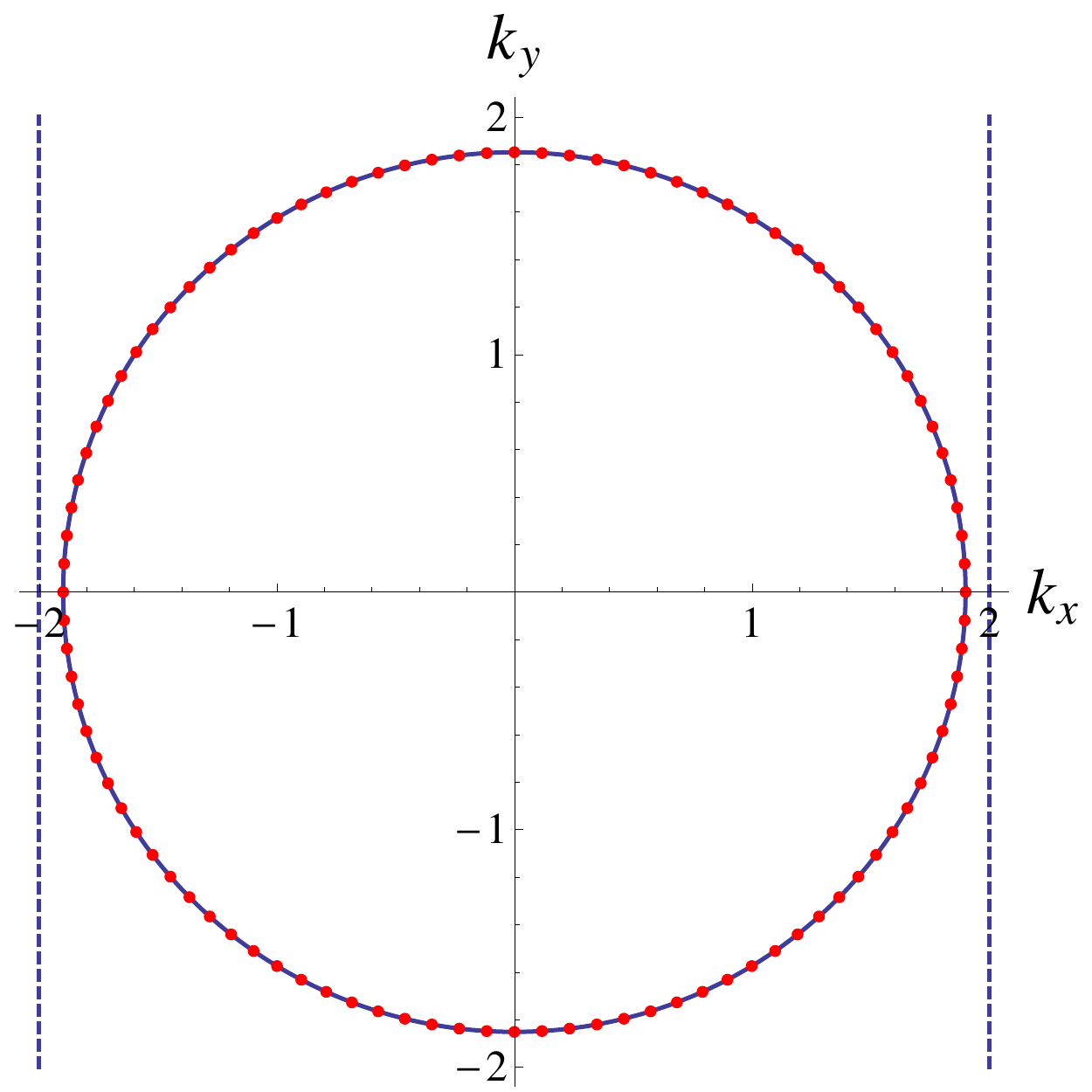}\hspace{0.1cm}
\includegraphics[scale=0.6]{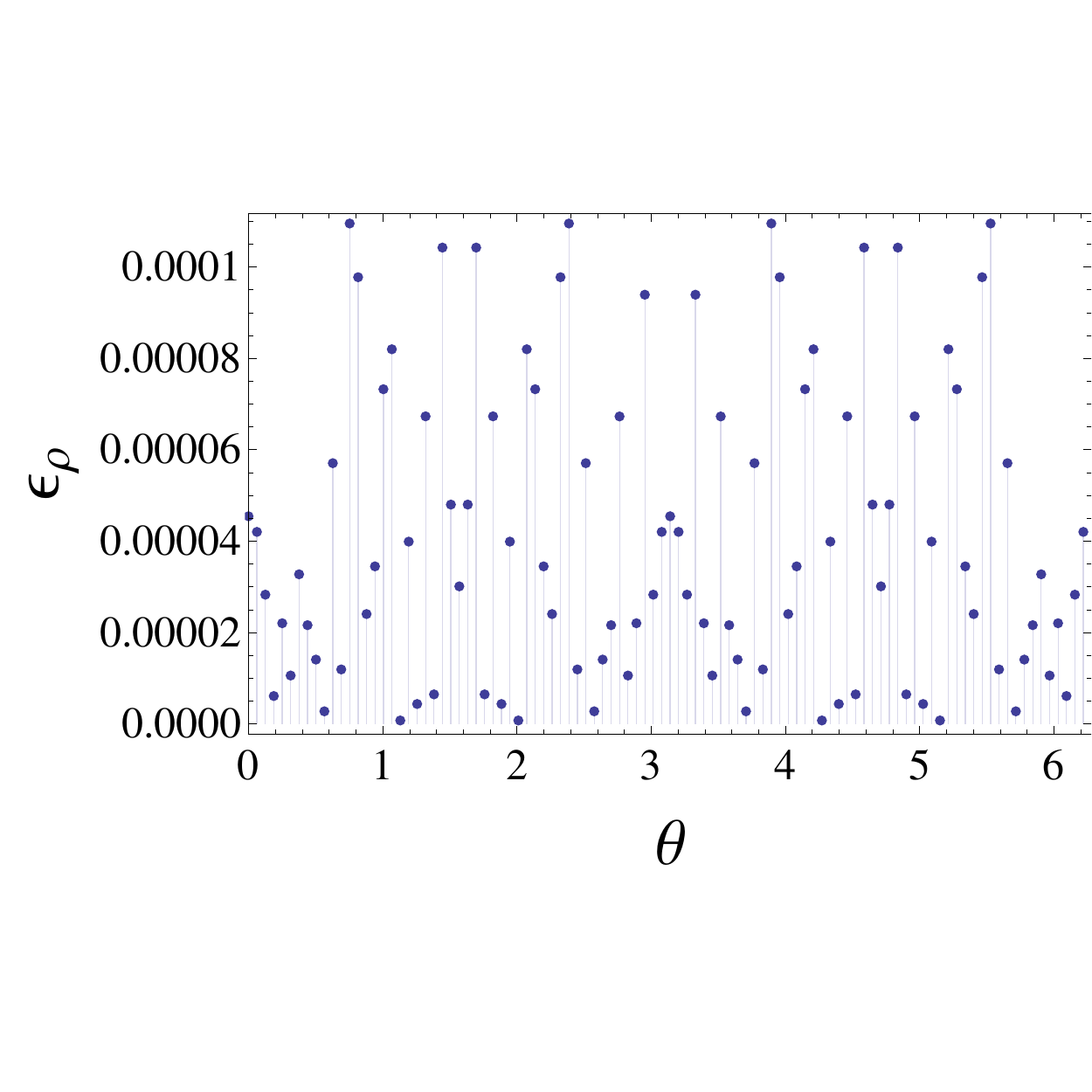}\\ \hspace{0.1cm}
\caption{\label{FS2}We show the shape of Fermi surface on the left
and the error bar fit by the ellipse in the polar coordinate on
the right for $k_0=2$, $A_0 =2$, $\mu = 2.35$, $T/\mu = 0.0081$
and $q=1.3$, where the Fermi momenta $k_F<2$, namely the Fermi
surface is located within the first Brillouin zone. } }
\end{figure}

With this in mind, we show in Fig.\ref{FS1} an example with the 3d
plot of $A(\omega,k)$ for the fixed $k_y=0$. A sharp peak occurs
near $\omega=0$, indicating a Fermi surface with the Fermi
momentum $k_F=2.634$. Similarly, we may locate the positions of
the Fermi surface for other values of $k_y$ such that the shape of
the Fermi surface can be plotted in the momentum space. We
illustrate our results in Fig.\ref{FS2} for the case of $k_0=2$,
$A_0 =2$, $\mu = 2.35$, $T/\mu = 0.0081$ and $q=1.3$, which
corresponds to a Fermi surface located within the first Brillouin
zone, namely $k_F< K/2=2$. Although the shape of Fermi surface
appears like a circle in the plot, our data clearly tells us that
it is not a circle any more. As a matter of fact, it can be
precisely fit by an equation of ellipse as follows
\begin{equation}
\frac{k_{x}^2}{a^2}+\frac{k_{y}^2}{b^2}=1
\end{equation}
with $a=1.8991$ and $b=1.8511$. We find this fitting has a very
high accuracy, which can be seen from our error bar analysis
presented on the right side of Fig.\ref{FS2}. Remarkably such a
result is universal in the sense that our elliptical Fermi surface
is robust against the values of parameters in question except that
the longer axis as well as the shorter axis is varied as it should
be. Now let us see how our elliptical Fermi surface varies with
our relevant parameters. As such, we introduce two quantities,
namely, the difference between the longer axis and shorter axis
$d=a-b$ as well as the eccentricity $e=\sqrt{a^2-b^2}/a$. The
relevant results are listed in Table \ref{amplitude},
\ref{temperature}, and \ref{charge}. Obviously, we observe the
following behaviors

\begin{itemize}
    \item As we increase the amplitude of our periodic source $A_0$, both $d$ and $e$ are increased.
        \item As we lower the temperature by increasing $\mu_1$, both $d$ and $e$ are increased.
    \item As we increase the charge $q$, the Fermi surface is enlarged with $e$ suppressed.
\end{itemize}

\begin{widetext}
\begin{table}[ht]
\begin{center}

\begin{tabular}{|c|c|c|c|c|c|c|}
         \hline
$~~A_0~~$ &~~$0.5$~~&~~$1$~~&~~$1.5$~~&~~$2$~~&~~$2.5$~~
          \\
        \hline
~~$d$~~ & ~~$0.004$~~ & ~~$0.013$~~ & ~~$0.029$~~ & ~~$0.048$~~ & ~~$0.069$~~
          \\
        \hline
~~$e$~~ & ~~$0.06414$~~ & ~~$0.1187$~~ & ~~$0.1742$~~ & ~~$0.2242$~~ & ~~$0.2675$~~
          \\
        \hline
\end{tabular}
\caption{\label{amplitude} The variation of $d$ and $e$ with the
amplitude $A_0$, where we have fixed the other parameters as
$k_0=2$, $\mu = 2.35$, $T/\mu = 0.0081$ and $q=1.3$.}

\end{center}
\end{table}
\end{widetext}

\begin{widetext}
\begin{table}[ht]
\begin{center}

\begin{tabular}{|c|c|c|c|c|c|c|}
         \hline
$~~\mu_1~~$ &~~$2$~~&~~$2.1$~~&~~$2.2$~~&~~$2.3$~~&~~$2.4$~~
          \\
        \hline
~~$d$~~ & ~~$0.019$~~ & ~~$0.021$~~ & ~~$0.024$~~ & ~~$0.027$~~ & ~~$0.031$~~
          \\
        \hline
~~$e$~~ & ~~$0.1487$~~ & ~~$0.1546$~~ & ~~$0.1610$~~ & ~~$0.1696$~~ & ~~$0.1792$~~
          \\
        \hline
\end{tabular}
\caption{\label{temperature} The variation of $d$ and $e$ with the
temperature which is controlled by $\mu_1$, where we have fixed the other parameters as
$k_0=2$, $A_0 =1.5$, $\mu = 2.35$ and $q=1.3$.}

\end{center}
\end{table}
\end{widetext}

\begin{widetext}
\begin{table}[ht]
\begin{center}

\begin{tabular}{|c|c|c|c|c|c|c|}
         \hline
$~~q~~$ &~~$0.5$~~&~~$0.7$~~&~~$0.9$~~&~~$1.1$~~&~~$1.3$~~
          \\
        \hline
~~$d$~~ & ~~$0.013$~~ & ~~$0.018$~~ & ~~$0.022$~~ & ~~$0.026$~~ & ~~$0.029$~~
          \\
        \hline
~~$e$~~ & ~~$0.2479$~~ & ~~$0.2258$~~ & ~~$0.2010$~~ &
~~$0.1871$~~ & ~~$0.1742$~~
          \\
        \hline
\end{tabular}
\caption{\label{charge} The variation of $d$ and $e$ with the
charge $q$, where we have fixed the other parameters as $k_0=2$,
$A_0 =1.5$ and $\mu =\mu_1=2.35$.}

\end{center}
\end{table}
\end{widetext}

\begin{figure}
\center{
\includegraphics[scale=0.4]{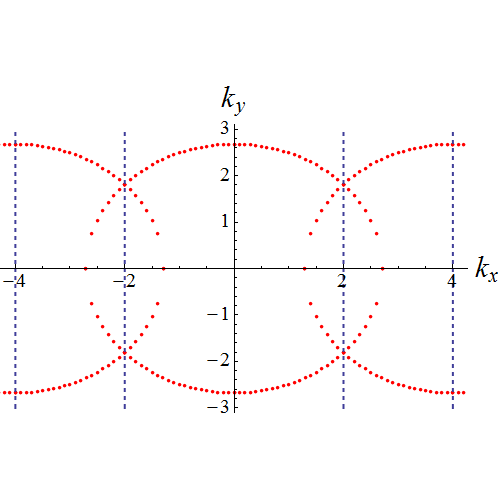}\hspace{0.1cm}
\caption{\label{gap1}Band structure for $q=1.7$,$k_0=2$, $A_0
=1.5$ and $\mu =\mu_1=2.35$($k_F>2$). } }
\end{figure}
\begin{figure}
\center{
\includegraphics[scale=0.4]{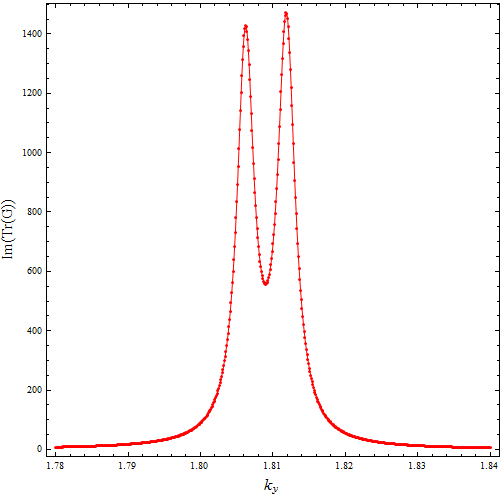}\hspace{0.1cm}
\includegraphics[scale=0.4]{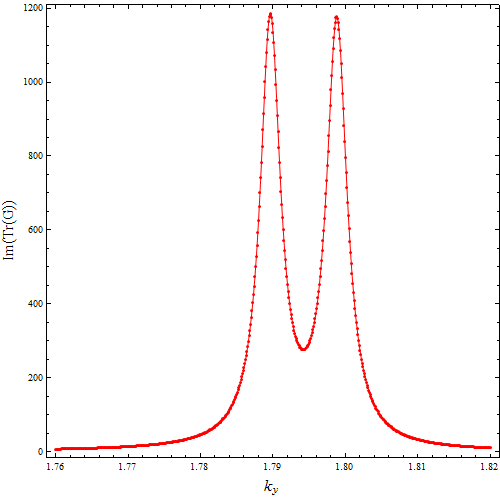}\\ \hspace{0.1cm}
\caption{\label{gap2} There are two peaks for the imaginary part
of the Green function at $k_x=2$(BZ boundary). The left one is for
$A_0=1.5$ while the right one for $A_0=2$, with $q=1.7$,$k_0=2$,
and $\mu =\mu_1=2.35$ fixed.  } }
\end{figure}

We remark that above phenomena can be observed in the ionic
lattice background as well. Interestingly, we find that in ionic
lattice case, the longer axis of the ellipse changes from $x$ to
$y$ direction.

\begin{figure}
\center{
\includegraphics[scale=0.4]{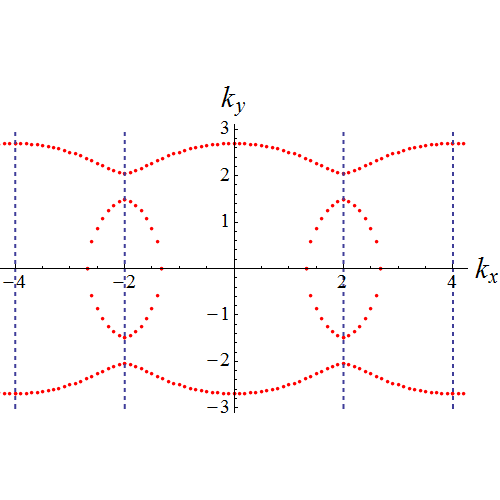}\hspace{0.1cm}
\includegraphics[scale=0.4]{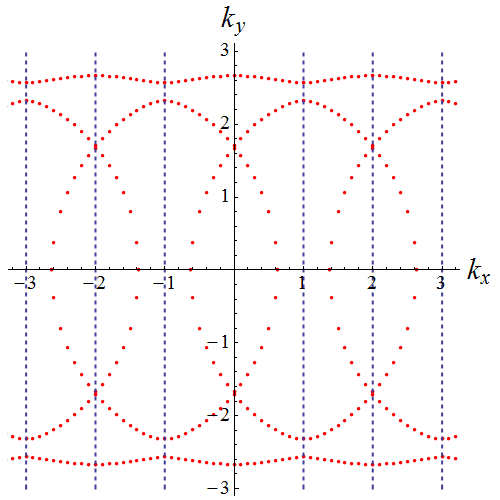}\\ \hspace{0.1cm}
\caption{\label{gap3}Band structure in ionic lattice model for
$k_F>K/2$$(q=1.7,A_0 =0.3,\mu =\mu_1=2.3)$ and $k_F>K$$(q=1.7,A_0
=0.1,\mu =\mu_1=2.3)$. Note that for the left one we have
 set $k_0=4$ such that $K=4$, while for the right one $k_0=2$ such that $K=2$.  } }
\end{figure}

When the charge $q$ is large enough, the Fermi surface will go
beyond the first Brillouin zone with $k_F>K/2$. Now let us turn to
such a situation by first demonstrating the band structure of
Fermi surface in Fig.\ref{gap1} for the scalar lattice. Note that
the Fermi surface exhibits a periodic structure along the $k_x$
direction as it should be the case guaranteed by Bloch theorem. On
the other hand, it appears that our Fermi surface does not show
the band gap structure at the intersection of the Fermi surface
with the Brillouin zone boundary, which is at odds with the
familiar lattice effect. In order to see if this is really the
case, we zoom in the Fermi surface precisely at the Brillouin zone
boundary. The corresponding result is plotted in Fig.\ref{gap2}.
To our pleasure, the band gap shows up and becomes large with the
increase of the amplitude of periodic source. So to see a clearer
band gap structure, it is better to move on to the ionic lattice
in which the lattice effects should be much stronger as we
discussed in previous section. As shown in Fig.\ref{gap3} for the
ionic lattice, the gap is evidently observed as one expects. From
this figure we also notice that the band structure becomes richer
when the Fermi surface intersects with more than two Brillouin
zones.

\section{Conclusion}
We have investigated the lattice effect on the Fermi surface by
putting the bulk Dirac field in the ultra cold holographic
lattices, where both the background equations and the probe Dirac
field are solved by pseudo-spectral method numerically. One
interesting result is that the Fermi surface is always modified by
the lattice from a circle to an ellipse. Suppose that such a Fermi
surface is formed by some kind of exotic free fermionic
quasi-particles, then the elliptical shape means that the
effective mass along the $x$ direction in scalar lattice (or the
$y$ direction in ionic lattice) becomes more massive by some kind
of renormalization effects due to the presence of the lattice.
In any case, such a universal behavior of holographic Fermi surface
begs for a deeper understanding. In addition, we holographically
reproduce the band gap structure at the intersection of the Fermi
surface and the Brillouin zone boundary, which is a well
known lattice effect on the Fermi surface in condensed matter
theory.

Note that such a band gap structure is also obtained in
\cite{Liu:2012tr}. We would like to compare
our work with \cite{Liu:2012tr}. The lattice considered in
that paper is a weak ionic lattice, where the back reaction
of the periodic chemical potential to the bulk geometry is
ignored. The advantage of such a weak potential limit is two fold.
One is that such a lattice can be put at zero temperature, and the
other is that the analytic technique developed in \cite{Faulkner:2009wj} can
be borrowed heavily to the relevant perturbation calculation. The
disadvantage is also obvious. First, the amplitude of spatially
varying chemical potential must be small enough, otherwise the
perturbation calculation will break down. Second, the neglect of
the back reaction may lead to the loss of some significant
physics. For example, the elliptical Fermi surface we find does
not appear in the perturbation calculation\footnote{Of course,
such a Fermi surface may show up when one goes to higher order in
the perturbation calculation.  After all, the perturbation
calculation is made only to second order in \cite{Liu:2012tr}.}. Third,
when such a perturbation calculation is made to higher order or
for the non-diagonal components of retarded Green function, the
analysis will become much involved. Compared with this, the full
retarded Green function can be obtained by our numerics once and
for all. In this sense, our paper and \cite{Liu:2012tr} are
complementary to each other.

We conclude with various issues worthy of further investigation.
First, even though it is numerically harder, it is apparently better to
construct the zero temperature holographic lattices for our probe
Dirac field to propagate in. Second, pertaining to the STM
experiment, it is important for us to extract the retarded Green
function in the position space, which can be achieved in the
following two ways. One is to Fourier transform our resultant
Green function in the momentum space to the position space. The
other is to work directly in the position space by imposing Dirac
delta source on the AdS boundary. We hope to address these issues
in the near future.
\begin{acknowledgments}
We are grateful to Liqing Fang, Xianhui Ge, Xiaomei Kuang, Yan
Liu, Yu Tian, Bin Wang, Shaofeng Wu and Xiaoning Wu for helpful
discussions. This work is supported by the Natural Science
Foundation of China under Grant Nos.11275208 and 11178002. Y.L.
also acknowledges the support from Jiangxi young scientists
(JingGang Star) program and 555 talent project of Jiangxi
Province. J.W. is also supported by the National Research
Foundation of Korea(NRF) grant funded by the Korea
government(MEST) through the Center for Quantum Spacetime(CQUeST)
of Sogang University with grant number 2005-0049409. H.Z. is
supported in part by the Belgian Federal Science Policy Office
through the Interuniversity Attraction Pole P7/37, by
FWO-Vlaanderen through the project G.0114.10N, and by the Vrije
Universiteit Brussel through the Strategic Research Program
"High-Energy Physics".

\end{acknowledgments}

\end{document}